\documentclass[aps,pre,preprint]{revtex4}
\usepackage[dvips]{epsfig}
\usepackage[dvips]{color}
\usepackage{graphicx}
\usepackage{float}
\usepackage{subfigure}
\usepackage[applemac]{inputenc}
\usepackage[T1]{fontenc}
\usepackage{amssymb}
\usepackage{amsmath}
\usepackage{bm}
\usepackage{mathrsfs}
\usepackage{float}
\usepackage{slashed}
\usepackage[english]{babel}

\begin{document}

\title{Effects of the electron spin on the nonlinear generation of
quasi-static magnetic fields in a plasma}
\author{M. STEFAN$^1$, G. BRODIN$^1$, F. HAAS$^{1, 2}$, and M. MARKLUND$^1$}
\affiliation{$^1$Department of Physics, Ume{\aa} University, SE--901 87 Ume{\aa}, Sweden \\
	$^2$Universidade do Vale do Rio dos Sinos, Av. Unisinos 950, 93022-000 S\~ao Leopoldo RS, Brazil}

\begin{abstract}
Through an extended kinetic model, we study the nonlinear generation of quasi-static magnetic fields by
high-frequency fields in a plasma, taking into account the effects of the
electron spin. It is found that although the largest part of the nonlinear
current in a moderate density, moderate temperature plasma is due to the
classical terms, the spin may still give a significant contribution to the
magnetic field generation mechanism. Applications of our results are
discussed. 
\end{abstract}


\maketitle

\section{Introduction}

The low-frequency fields nonlinearly generated by high-frequency waves in a
plasma has been the subject of much studies during several decades \cite%
{Washimi,kono,Stenflo,Belkov,Jovanovic,Prabhakar2,Srivastava,Aliev1,Li,Zhu,Brodin,Brodin-rel,Liu and Li,Lazar2006-1,Lazar2006}%
. The focus has been on the ponderomotive force, see e.g. \cite{Washimi,kono}%
, on heating nonlinearities due to collisional effects \cite{Stenflo}, or on
the possibility to generate quasi-static magnetic fields \cite%
{Belkov,Jovanovic,Prabhakar2,Srivastava,Aliev1,Li,Zhu,Brodin,Brodin-rel,Liu
and Li,Lazar2006-1,Lazar2006}. Generation of such fields has proven to be
rather sensitive to small deviations from the classical collisionfree Vlasov
model. Thus Ref. \cite{Brodin} found that a small collision frequency
(described by the Lorenz collision model) was sufficient to significantly
alter the predictions from the Vlasov equation. Furthermore, Refs \cite%
{Brodin-rel,Lazar2006} found that weak relativistic effects (with the
thermal energy much smaller than the electron rest mass energy), also could
alter the nonlinear generation of quasi-static magnetic fields.

In the present paper we will investigate to what extent quantum mechanical
effects \cite{Quantum
plasmas,Manfredi,Haas,Tito-2008a,Tito-2008b,Tito-Rydberg-plasma,Marklund-EPL,Classical-quant,Spin-ref}%
, and in particular electron spin effects \cite%
{Classical-quant,Spin-ref,soliton1,soliton2,Ferro-PRE,cowley,g-factor,Full-quantum}, can
change the nonlinear low-frequency response to high-frequency fields. A
semi-classical model for spin effects generalizing the Vlasov equation was
presented decades ago \cite{cowley}. Recently it was improved to include the
magnetic dipole force and the magnetization current due to the spin \cite%
{g-factor}, and finally a fully quantum mechanical model was given \cite%
{Full-quantum}, reducing to the Wigner equation \cite{Manfredi} without spin
effects, which can be further reduced to the classical Vlasov equation. Here
we start from the long wavelength limit (spatial scales much longer than the
thermal de Broglie wavelength) of the full theory \cite{Full-quantum}, where
all quantum mechanical effects are directly associated with the spin. A
formal expression for the nonlinear low-frequency current is defined for a
general geometry. This expression is then evaluated for two special cases
with specified wave-vectors and polarization of the high-frequency fields.
It is then found that although the classical terms gives the largest
contribution to the nonlinear current for a plasma of moderate density and
temperature, the spin terms can still contribute significantly to the
generation of quasi-static magnetic fields.

The organization of the paper is as follows. In Section II the calculation
procedure starting from the classical Vlasov equation is outlined, and the
nonlinear low-frequency current corresponding to this case is presented.\ In
Section III the spin kinetic equation is introduced, and our main results
starting from this model is derived. In Section IV we investigate the
generation of quasi-static magnetic fields by the nonlinear current
densities that we have derived. In section V, our results summarized and
discussed. Finally, the Appendix addresses the possible significance of
quantum mechanical effects left out in the previous calculations.

\section{Classical kinetic model}

In order to outline our method of calculating the nonlinear current, we
shall first consider the nonlinear mixing of two high-frequency waves $%
(\omega _{1},\mathbf{k}_{1})$ and $(\omega _{2},\mathbf{k}_{2})$ in a
collisionless plasma. Considering only electron motion we describe the
evolution of the perturbation $f$ of the electron distribution function by
means of the classical Vlasov equation 
\begin{equation}
\frac{\partial f}{\partial t}+\mathbf{v}\cdot \nabla f+\frac{q}{m}(\mathbf{E}%
+\mathbf{v}\times \mathbf{B})\cdot \frac{\partial F_{0}}{\partial \mathbf{v}}%
=-\frac{q}{m}(\mathbf{E}+\mathbf{v}\times \mathbf{B})\cdot \frac{\partial f}{%
\partial \mathbf{v}}  \label{Class-vlasov}
\end{equation}%
where $q/m$ is the electron charge to mass ratio and $F_{0}(\mathbf{v})$ is
the unperturbed velocity distribution function normalized such that $%
n_{0}=\int F_{0}d^{3}v$. For simplicity we assume that $F_{0}$ is
Maxwellian, i.e. $F_{0}\sim \exp (-v^{2}/2v_{t}^{2})$. Relativistic effects
are neglected. For notational convenience we assume that $\omega _{2}$ is
negative. The nonlinear mixing between the two high-frequency waves thus
yields a low-frequency response at $(\omega ,\mathbf{k})$ where $\omega
=\omega _{1}+\omega _{2}$ and $\mathbf{k}=\mathbf{k}_{1}+\mathbf{k}_{2}$.
Replacing the right hand side of equation (\ref{Class-vlasov}) by $(\mathbf{E%
}_{1}+\mathbf{v}\times \mathbf{B}_{1})\cdot (\partial f_{2}/\partial \mathbf{%
v})+(1\leftrightarrow 2)$ where $(1\leftrightarrow 2)$ means interchange of
indices 1 and 2, and $f_{2}\approx -(iq/m)\mathbf{E}_{2}\cdot (\partial
F_{0}/\partial \mathbf{v})/(\omega _{2}-\mathbf{k}_{2}\cdot \mathbf{v})$, we
then obtain the nonlinear and classical part of the generated low-frequency
current density ${\mathbf{J}}_{\mathrm{cl}}$ as 
\begin{equation}
{\mathbf{J}}_{\mathrm{cl}}=-\frac{iq^{2}}{m}\int \frac{\mathbf{v}}{(\omega -%
\mathbf{k}\cdot \mathbf{v})}\left[ (\mathbf{E}_{1}+\mathbf{v}\times \mathbf{B%
}_{1})\cdot \frac{\partial f_{2}}{\partial \mathbf{v}}+(1\leftrightarrow 2)%
\right] d\mathbf{v}
\end{equation}%
which can be directly rewritten in the form 
\begin{eqnarray}
{\mathbf{J}}_{\mathrm{cl}} &=&\frac{q^{3}}{m^{2}v_{t}^{2}}\int \frac{\mathbf{%
v}F_{0}}{\omega -\mathbf{k}\cdot \mathbf{v}}\left[ \frac{(\mathbf{E}_{1}+%
\mathbf{v}\times \mathbf{B}_{1})\cdot \mathbf{E}_{2}}{(\omega _{2}-\mathbf{k}%
_{2}\cdot \mathbf{v})}-\right.  \notag \\
&&\left. \frac{\mathbf{E}_{1}\cdot \mathbf{v}\mathbf{E}_{2}\cdot \mathbf{v}}{%
v_{t}^{2}(\omega _{2}-\mathbf{k}_{2}\cdot \mathbf{v})}+\frac{\mathbf{k}%
_{2}\cdot (\mathbf{E}_{1}+\mathbf{v}\times \mathbf{B}_{1})\mathbf{E}%
_{2}\cdot \mathbf{v}}{(\omega _{2}-\mathbf{k}_{2}\cdot \mathbf{v})^{2}}%
+(1\leftrightarrow 2)\right] d^{3}v
\end{eqnarray}%
Using the fact that the phase velocities of the high-frequency waves are
much larger than the thermal velocity $v_{t}$ we can adopt $\mathbf{k}%
_{1,2}\cdot \mathbf{v}/\omega _{1,2}$ and $\omega /\omega _{1,2}$ as
expansion parameters and obtain, up to order $(\omega /\omega _{1,2})^{3}$ 
\begin{eqnarray}
{\mathbf{J}}_{\mathrm{cl}} &=&\frac{q^{3}}{m^{2}v_{t}^{2}\omega _{1}\omega
_{2}}\int \mathbf{v}F_{0}\left\{ \frac{\omega -2\mathbf{k}\cdot \mathbf{v}}{%
\omega -\mathbf{k}\cdot \mathbf{v}}\left[ \mathbf{E}_{1}\cdot \mathbf{E}%
_{2}\left( 1+\frac{\mathbf{k}_{1}\cdot \mathbf{v}}{\omega _{1}}+\frac{%
\mathbf{k}_{2}\cdot \mathbf{v}}{\omega _{2}}\right) +\right. \right.  \notag
\\
&&\left. \left. \mathbf{v}\cdot \mathbf{E}_{1}\frac{\mathbf{k}_{1}\cdot 
\mathbf{E}_{2}}{\omega _{1}}+\mathbf{v}\cdot \mathbf{E}_{2}\frac{\mathbf{k}%
_{2}\cdot \mathbf{E}_{1}}{\omega _{2}}\right] -\frac{\mathbf{v}\cdot \mathbf{%
E}_{1}\mathbf{v}\cdot \mathbf{E}_{2}}{v_{t}^{2}}\left( \frac{\mathbf{k}%
_{1}\cdot \mathbf{v}}{\omega _{1}}+\frac{\mathbf{k}_{2}\cdot \mathbf{v}}{%
\omega _{2}}\right) \right\} d^{3}v  \label{current-expanded}
\end{eqnarray}%
The first term of Eq. (\ref{current-expanded}) (proportional to $\mathbf{E}%
_{1}\cdot \mathbf{E}_{2}$) is one order larger in the expansion made, but
other terms are important as they are the ones contributing the generation
of quasi-static magnetic fields.

\section{Spin kinetic model}

Next we consider an extended distribution function $f(\mathbf{r},\mathbf{v},%
\mathbf{s},t)$, where $\mathbf{s}$ is a spin vector with fixed length,
describing the orientation of the electron spin. On semi-classical grounds
it can be shown \cite{g-factor} that $df/dt=0$, where $d/dt$ is a total time
derivative following the generalized particle orbit, i.e. including the spin
evolution. Using the Heisenberg equation of motion to get $d\mathbf{s}/dt$,
a semi-classical model including the spin degree of freedom is found \cite%
{g-factor}. With certain corrections, such a model can be derived as the
long scale limit (scale lengths much longer than the thermal de Broglie
wavelength) of a fully quantum mechanical treatment \cite{Full-quantum}. The
evolution equation in this regime is then found to be

\begin{equation}
\partial _{t}f+\mathbf{v}\cdot \nabla f+\left[ \frac{q}{m}\left( \mathbf{E}+%
\mathbf{v}\times \mathbf{B}\right) +\frac{2\mu _{e}}{m}\nabla \left( \mathbf{%
s}\cdot \mathbf{B}\right) +\frac{3\mu _{e}}{2m}\nabla \left( \mathbf{B}\cdot
\nabla _{\mathbf{s}}\right) \right] \cdot \nabla _{\mathbf{v}}f+\frac{2\mu
_{e}}{\hbar }\left( \mathbf{s}\times \mathbf{B}\right) \cdot \nabla _{s}f=0
\label{Gen-Vlasov-spin}
\end{equation}%
where we here use the convention that $\left\vert \mathbf{s}\right\vert =1$.
The difference from the semi-classical theory deduced from $df/dt=0$ is
twofold. Firstly, the constant coefficients of the spin terms differ
somewhat (this is compensated by different coefficients in the spin current
given below, such that also the semi-classical model is energy conserving).
Secondly, the term proportional to $\mathbf{B}\cdot \nabla _{\mathbf{s}}$
has no semi-classical counterpart. This term can be viewed as part of the
magnetic dipole force, which account for the fact that a quantum mechanical
probability distribution of a single particle spin is always smeared out, as
opposed to a classical spin (or magnetic dipole moment) which has a unique
direction. The spin kinetic model is completed by Maxwell's equations with
the current density 
\begin{equation}
\mathbf{J}=\mathbf{J}_{f}+\mathbf{J}_{m}=q\int \mathbf{v}f\!d^{3}v\,d\Omega
_{s}+\mu _{e}\int \mathbf{s}f\!d^{3}v\,d\Omega _{s}  \label{current-density}
\end{equation}%
where $\mathbf{J}_{f}$ and $\mathbf{J}_{m}$ is the free part and
magnetization (spin) part of the current density, respectively. Using
spherical coordinates for the spin angles we have $d\Omega _{s}=\sin \theta
_{s}d\theta _{s}d\varphi _{s}$.

Next we follow the steps of section II, and write the nonlinear
high-frequency source terms (due to $(\omega _{1},\mathbf{k}_{1})$ and $%
(\omega _{2},\mathbf{k}_{2})$) on the right hand side. Since the zero order
distribution function $F_{0}$ is isotropic and independent of $\mathbf{s}$
we find 
\begin{align}
\partial _{t}f+\mathbf{v}\cdot \nabla f+\left[ \frac{q}{m}\mathbf{E}+\frac{%
2\mu _{e}}{m\hbar }\nabla \left( \mathbf{s}\cdot \mathbf{B}\right) \right]
\cdot \nabla _{\mathbf{v}}F_{0}=& -\left[ \frac{q}{m}\left( \mathbf{E}+%
\mathbf{v}\times \mathbf{B}\right) +\frac{2\mu _{e}}{m\hbar }\nabla \left( 
\mathbf{s}\cdot \mathbf{B}\right) -\frac{3\mu _{e}\hbar }{2m}\nabla \left( 
\mathbf{B}\cdot \nabla _{\mathbf{s}}\right) \right] \cdot \nabla _{\mathbf{v}%
}f  \notag \\
& -\frac{2\mu _{e}}{\hbar }\left( \mathbf{s}\times \mathbf{B}\right) \cdot
\nabla _{s}f.  \label{vlasov}
\end{align}%
Following the same procedure as in the classical case we consider a
Maxwellian background distribution and use the notation 
\begin{align}
\omega & =\omega _{1}+\omega _{2}, \\
\mathbf{k}& =\mathbf{k}_{1}+\mathbf{k}_{2}.
\end{align}%
Since the zero order distribution function is isotropic by assumption, the
first order distribution function can be approximated as 
\begin{equation}
f_{1,2}\approx \frac{-i\left( q\mathbf{E}_{1,2}+\frac{2\mu _{e}}{\hbar }%
\nabla (\mathbf{s}\cdot \mathbf{B}_{1,2})\right) \cdot \partial _{\mathbf{v}%
}F_{0}}{m(\omega _{1,2}-\mathbf{k}_{1,2}\cdot \mathbf{v})},  \label{f1}
\end{equation}%
and we get the free nonlinear current density 
\begin{align}
\mathbf{J}_{\mathrm{f}}=-iq\int \!d^{3}v\,d\Omega _{s}\,\frac{\mathbf{v}}{%
\omega -\mathbf{k}\cdot \mathbf{v}}& \left\{ \left[ \frac{q}{m}\left( 
\mathbf{E}_{1}+\mathbf{v}\times \mathbf{B}_{1}\right) +\frac{2\mu _{e}}{%
m\hbar }\nabla \left( \mathbf{s}\cdot \mathbf{B}\right) +\frac{3\mu
_{e}\hbar }{2m}\nabla \left( \mathbf{B}\cdot \nabla _{\mathbf{s}}\right) %
\right] \cdot \partial _{\mathbf{v}}f_{2}\right.  \notag \\
& \left. +\frac{2\mu _{e}}{\hbar }\left( \mathbf{s}\times \mathbf{B}%
_{1}\right) \cdot \partial _{\mathbf{s}}f_{2}+\left( 1\leftrightarrow
2\right) \right\} ,  \label{freecurrent}
\end{align}%
and the nonlinear magnetization current 
\begin{align}
\mathbf{J}_{\mathrm{m}}=-i\nabla \times \int \!d^{3}v\,d\Omega _{s}\,\frac{%
\mathbf{s}}{\omega -\mathbf{k}\cdot \mathbf{v}}& \left\{ \left[ \frac{q}{m}%
\left( \mathbf{E}_{1}+\mathbf{v}\times \mathbf{B}_{1}\right) +\frac{2\mu _{e}%
}{m\hbar }\nabla \left( \mathbf{s}\cdot \mathbf{B}\right) +\frac{3\mu
_{e}\hbar }{2m}\nabla \left( \mathbf{B}\cdot \nabla _{\mathbf{s}}\right) %
\right] \cdot \partial _{\mathbf{v}}f_{2}\right.  \notag \\
& \left. +\frac{2\mu _{e}}{\hbar }\left( \mathbf{s}\times \mathbf{B}%
_{1}\right) \cdot \partial _{\mathbf{s}}f_{2}+\left( 1\leftrightarrow
2\right) \right\} .  \label{magcurrent}
\end{align}

A general geometry with arbitrary polarizations and wavevectors of the
high-frequency waves leads to extremely lengthy algebra, and is not within
the scope of the present paper. Thus as our first special case we consider
parallel polarization and propagation. For definiteness we choose the
geometry 
\begin{align}
\mathbf{k}_{1,2}& =k_{1,2}\mathbf{\hat{z}}, \\
\mathbf{B}_{1,2}& =B_{1,2}\mathbf{\hat{y}}, \\
\mathbf{E}_{1,2}& =E_{1,2}\mathbf{\hat{x}}.
\end{align}%
The high-frequency waves may be co-propagating or counter-propagating
depending on the sign of $k_{1,2}$. In this case subtracting the classical
contribution ${\mathbf{J}}_{\mathrm{cl}}$ we obtain that the free part and
the magnetization part of the nonlinear current due to the spin, denoted ${%
\mathbf{J}}_{\mathrm{sp}}$ combines to 
\begin{align}
{\mathbf{J}}_{\mathrm{sp}}=& -q\frac{3\mu _{e}^{2}}{m^{2}v_{t}^{2}}%
k_{1}k_{2}B_{1}B_{2}\int \!d^{3}v\,d\Omega _{s}\,\frac{\mathbf{v}}{\omega -%
\mathbf{k}\cdot \mathbf{v}}F_{0}  \notag \\
& \times \left[ \frac{1}{(\omega _{2}-k_{2}v_{z})}\left( 1-\frac{v_{z}^{2}}{%
v_{t}^{2}}+\frac{k_{2}v_{z}}{\omega _{2}-k_{2}v_{z}}\right) +\frac{1}{%
(\omega _{1}-k_{1}v_{z})}\left( 1-\frac{v_{z}^{2}}{v_{t}^{2}}+\frac{%
k_{1}v_{z}}{\omega _{1}-k_{1}v_{z}}\right) \right]
\end{align}%
which for $v_t \ll \omega_1/k_1 , \omega_2 / k_2$ reduces to 
\begin{equation}
{\mathbf{J}}_{\mathrm{sp}}=-q\frac{3\mu _{e}^{2}}{m^{2}v_{t}^{2}}%
k_{1}k_{2}B_{1}B_{2}\int \!d^{3}v\,d\Omega _{s}\,\frac{\mathbf{v}}{\omega -%
\mathbf{k}\cdot \mathbf{v}}F_{0}\left[ \frac{1}{\omega _{2}}\left( 1-\frac{%
v_{z}^{2}}{v_{t}^{2}}+\frac{k_{2}v_{z}}{\omega _{2}}\right) +\frac{1}{\omega
_{1}}\left( 1-\frac{v_{z}^{2}}{v_{t}^{2}}+\frac{k_{1}v_{z}}{\omega _{1}}%
\right) \right] .  \label{par-pol.prop}
\end{equation}%
Since $\mathbf{k}$ is along $\mathbf{\hat{z}}$, we note that ${\mathbf{J}}_{%
\mathrm{sp}}$ in (\ref{par-pol.prop}) will be in the $\mathbf{\hat{z}}$%
-direction. This is the same direction as the classical contribution found
from (\ref{current-expanded}). Furthermore, for moderate temperatures and
densities the classical contribution will be larger than that due to the
spin, and hence the result given by (\ref{par-pol.prop}) is merely a small
correction, at least if we limit ourselves to parameters of temperature and
density corresponding to laboratory and space plasmas.

Next we modify our special case and consider the polarization of the waves
to be perpendicular. As it turns out, the spin contribution to the nonlinear
current then vanishes, although there will be a significant magnetization in
the $\mathbf{\hat{z}}$-direction. If we also modify the wavevectors
slightly, this nonzero magnetization contributes to a nonlinear current
density. This means that we consider the following geometry: 
\begin{align}
\mathbf{E}_{1}& =E_{1}\mathbf{\hat{x}}, \\
\mathbf{E}_{2}& =-E_{2}\mathbf{\hat{y}}, \\
\mathbf{k}_{1}& =k_{1}\mathbf{\hat{z}}, \\
\mathbf{k}_{2}& =k_{2}\mathbf{\hat{z}}+\Delta k_{2}\mathbf{\hat{x}}, \\
\mathbf{B}_{1}& =B_{1}\mathbf{\hat{y}}, \\
\mathbf{B}_{2}& =B_{2}\mathbf{\hat{x}}+\Delta B_{2}\mathbf{\hat{z}}.
\end{align}%
To limit the algebra, the deviation from parallel propagation is considered
to be small, and we will thus only consider terms to first order in $\Delta
k_{2}$ or $\Delta B_{2}$. By Faradays law we note that $\Delta
k_{2}/k_{2}=\Delta B_{2}/B_{2}$. When we calculate the nonlinear currents in
this geometry it turns out that the free part of the current vanishes, i.e. $%
\mathbf{J}_{\mathrm{sp}}=\mathbf{J}_{\mathrm{sp-m}}$ and the nonlinear spin
current density reduces to 
\begin{equation}
\mathbf{J}_{\mathrm{sp}}=-\mathbf{\hat{y}}\frac{16\mu _{e}^{3}}{%
mv_{t}^{2}\hbar }\pi \,\Delta k_{2}\int \!d^{3}v\,B_{1}B_{2}\frac{F_{0}v_{z}%
}{\omega -\mathbf{k}\cdot \mathbf{v}}\left[ \frac{k_{1}\left( \frac{k}{k_{2}}%
+1\right) }{\omega _{1}-\mathbf{k}_{1}\cdot \mathbf{v}}+\frac{k_{2}\left( 
\frac{k}{k_{2}}-1\right) }{\omega _{2}-\mathbf{k}_{2}\cdot \mathbf{v}}\right]
\label{spin-ortho}
\end{equation}%
which vanish if we let $\Delta k_{2}\rightarrow 0$. Taking the limit of low
temperature, i.e $v_t \ll \omega / k$, this further simplifies to 
\begin{equation}
\mathbf{J}_{\mathrm{sp}}=-\mathbf{\hat{y}}\frac{16\mu _{e}^{3}}{m\hbar }\pi
\,\Delta k_{2}n_{0}\frac{k}{\omega ^{2}}\left[ \frac{k_{1}}{\omega _{1}}%
\left( \frac{k}{k_{2}}+1\right) +\frac{k_{2}}{\omega _{2}}\left( \frac{k}{%
k_{2}}-1\right) \right] B_{1}B_{2}  \label{spin-ortho-lt}
\end{equation}%
The geometry in this special case is of more interest, as we will find that
the spin-contribution can give a larger contribution to the generation of
quasi-static magnetic fields than the free current, also for relatively
modest plasma temperatures and densities. This issue will be investigated
further in section IV.

\subsection{A general isotropic background distribution}

The above results can easily be generalized to be valid for any background
distribution that is a function of $v^{2}$, i.e. isotropic, which is of
interest e.g. for a dense plasma when the thermodynamic equilibrium
distribution is a Fermi-Dirac distribution rather than a Maxwellian. With
this more general background distribution we get in the case of parallel
polarization the nonlinear spin current 
\begin{align}
\mathbf{J}_{\mathrm{sp}}=& q\frac{6\mu _{e}^{2}}{m^{2}}k_{1}k_{2}B_{1}B_{2}%
\int \!d^{3}v\,d\Omega _{s}\,\frac{\mathbf{v}}{\omega -\mathbf{k}\cdot 
\mathbf{v}}  \notag \\
& \times \left[ \frac{1}{(\omega _{2}-k_{2}v_{z})}\left( 1-\frac{v_{z}^{2}}{%
v_{t}^{2}}+\frac{k_{2}v_{z}}{\omega _{2}-k_{2}v_{z}}\right) +\frac{1}{%
(\omega _{1}-k_{1}v_{z})}\left( 1-\frac{v_{z}^{2}}{v_{t}^{2}}+\frac{%
k_{1}v_{z}}{\omega _{1}-k_{1}v_{z}}\right) \right] \frac{\partial }{\partial
(v^{2})}F_{0},
\end{align}%
reducing to 
\begin{equation}
\mathbf{J}_{\mathrm{sp}}=q\frac{6\mu _{e}^{2}}{m^{2}}k_{1}k_{2}B_{1}B_{2}%
\int \!d^{3}v\,d\Omega _{s}\,\frac{\mathbf{v}}{\omega -\mathbf{k}\cdot 
\mathbf{v}}\left[ \frac{1}{\omega _{2}}\left( 1-\frac{v_{z}^{2}}{v_{t}^{2}}+%
\frac{k_{2}v_{z}}{\omega _{2}}\right) +\frac{1}{\omega _{1}}\left( 1-\frac{%
v_{z}^{2}}{v_{t}^{2}}+\frac{k_{1}v_{z}}{\omega _{1}}\right) \right] \frac{%
\partial }{\partial (v^{2})}F_{0}
\end{equation}%
for the limit $v_t \ll \omega_1 / k_1, \omega_2 / k_2$.

In the case of orthogonal polarization in the same way we obtain 
\begin{equation}
\mathbf{J}_{\mathrm{sp}}=\mathbf{\hat{y}}\frac{32\mu _{e}^{3}}{m\hbar }\pi
\,\Delta k_{2}\int \!d^{3}v\,B_{1}B_{2}\frac{v_{z}}{\omega -\mathbf{k}\cdot 
\mathbf{v}}\left[ \frac{k_{1}\left( \frac{k}{k_{2}}+1\right) }{\omega _{1}-%
\mathbf{k}_{1}\cdot \mathbf{v}}+\frac{k_{2}\left( \frac{k}{k_{2}}-1\right) }{%
\omega _{2}-\mathbf{k}_{2}\cdot \mathbf{v}}\right] \frac{\partial }{\partial
(v^{2})}F_{0}
\end{equation}%
and in the low temperature limit this reduces to (\ref{spin-ortho-lt}).

\section{Magnetic field generation}

To demonstrate the significance of the spin contribution in the nonlinear
current density we write Ampere's law as 
\begin{equation}
i\mathbf{k\times B(}\omega \mathbf{,k)=}\mu _{0}\left( \mathbf{J}_{\mathrm{l}%
}\mathbf{(}\omega \mathbf{,k)+J}_{\mathrm{nl}}\mathbf{(}\omega \mathbf{,k)}%
\right) -\frac{i\omega }{c^{2}}\mathbf{E(}\omega \mathbf{,k).}
\label{Ampere}
\end{equation}%
Next we write the linear current density as $\mathbf{J}_{\mathrm{l}}\mathbf{(%
}\omega \mathbf{,k)=}\sigma \mathbf{(}\omega \mathbf{,k)E(}\omega \mathbf{,k)%
}$, where $\sigma \mathbf{(}\omega \mathbf{,k)}$ is the linear conductivity.
Combining (\ref{Ampere}) with Faraday's law, we immediately find the low
frequency magnetic field generated by the nonlinear current as 
\begin{equation}
\mathbf{B(}\omega \mathbf{,k)}=\frac{i\mu _{0}\mathbf{k\times J}_{\mathrm{nl}%
}}{D_{\mathrm{em}}\mathbf{(}\omega \mathbf{,k)}}=\frac{i\mu _{0}\mathbf{%
k\times (J}_{\mathrm{cl}}+\mathbf{J}_{\mathrm{sp}})}{D_{\mathrm{em}}\mathbf{(%
}\omega \mathbf{,k)}}  \label{Magn-gen}
\end{equation}%
where $D_{\mathrm{em}}\mathbf{(}\omega \mathbf{,k)}=\omega
^{2}-k^{2}c^{2}-i\omega \sigma \mathbf{(}\omega \mathbf{,k)\simeq }\omega
^{2}-k^{2}c^{2}-\omega _{p}^{2}$. \ Eq. (\ref{Magn-gen}) thus contains both
the classical contribution from (\ref{current-expanded}) as well as the spin
contribution from (\ref{spin-ortho}). To shed further light on this
expression we evaluate (\ref{Magn-gen}) in the low temperature regime $%
kv_{th}\ll \omega $, in which case the expression simplifies to 
\begin{equation}
\mathbf{B(}\omega \mathbf{,k)}=-\frac{i\mu _{0}k}{D_{em}(\omega ,\bm k)}n_{0}%
\frac{B_{1}B_{2}}{m}\frac{\Delta k_{2}}{\omega _{2}}\left\{ \frac{%
q^{3}v_{t}^{2}(k_{1}^{2}+k_{2}^{2})}{mk_{1}k_{2}\omega _{2}}+\frac{16\mu
_{e}^{3}\pi k_{2}}{\hbar }\left[ \frac{k_{1}}{\omega _{1}}\left( \frac{k}{%
k_{2}}+1\right) +\frac{k_{2}}{\omega _{2}}\left( \frac{k}{k_{2}}-1\right) %
\right] \right\}  \label{class-spin-comp}
\end{equation}%
Thus we see that the spin contribution to the magnetic field generation \
(proportional to $\mu _{e}^{3}/\hbar $) dominates in the regime 
\begin{equation}
\hbar ^{2}k_{1}k_{2}\gtrsim \frac{m^{2}v_{th}^{2}}{2}  \label{simple-cond}
\end{equation}%
whereas the classical contribution dominates otherwise. A case of
experimental interest clearly includes two high-frequency sources. However,
we can note that our results also are of relevance for a single source,
where $(\omega _{1},\mathbf{k}_{1})$ and $(\omega _{2},\mathbf{k}_{2})$
represents different spectral components of a focused pulse. In this case
typically $\left\vert \mathbf{k}\right\vert \ll \left\vert \mathbf{k}%
_{1,2}\right\vert $. For the condition (\ref{simple-cond}) it does not
matter whether two pulses or a single source is used. Furthermore, for
current laser plasma experiments with lasers in the optical regime, it is
clear from (\ref{simple-cond}) that the classical contribution will
dominate. However, in case laser-plasma experiments with an X-Fel source
such as that being built in DESY is made \cite{Desy}, $\left\vert \mathbf{k}%
_{1,2}\right\vert \sim 6\times 10^{9}\mathrm{m}^{-1}$, and the quantum spin
effects can be of importance for magnetic field generation for plasma
temperatures $T\lesssim 2$ $\times 10^{4}\mathrm{K}$.This condition have
been derived for the case of a specified geometry, and it should be noted
that the results may differ in case polarizations and/or the directions of
wave vectors are changed.

\section{Summary and discussion}

In the present paper we have studied the nonlinear current density generated
by high-frequency waves in a plasma, with a focus on the contribution
emanating from the electron spin. The largest part of the current density is
usually associated with the classical ponderomotive effect. However, it is
found that although the largest part of the nonlinear current in a moderate
density, moderate temperature plasma is due to the classical terms, the spin
may still give a significant contribution to the magnetic field generation
mechanism. For the geometry considered here, the condition needed for spin
effects to be important require short-wavelength sources, of the order of
the x-ray regime. Besides the quantum effects due to spin considered here,
there is also particle dispersive quantum effects. Although a thorough
consideration of such effects is still to be made, our calculations outlines
here indicate that particle dispersive terms may be of comparable importance
for the nonlinear current density. Thus there still remains much research in
this area to be made.

\acknowledgments
MM was supported by the Swedish Research Council Contract \# 2007-4422 and the European Research Council Contract \# 204059-QPQV.

\appendix*

\section{Particle dispersive effects}

The quantum effects due to spin that we have considered here should be
compared to the ones due to particle dispersive effects. Such effects can be
described by the Wigner function, that reduces to the classical distribution
function whenever the thermal de Broglie wavelength is small compared to the
scale lengths of the problem. The quantum corrections due to this in the
kinetic equation scale as $\hbar^2$, whereas e.g. the magnetic dipole force
due to the spin scale as $\hbar$. However, this does not necessarily mean
that the lowest order quantum corrections is due to the spin, since for a
spin independent zero order distribution function, the spin term at one
place (e.g. the magnetic dipole force) always need to be combined with
another spin effect (e.g. the magnetization current) to produce a
non-vanishing nonlinear current. Thus both the spin effects and particle
dispersive effects produce quantum corrections that are proportional to $%
\hbar ^{2}$ to lowest order. While our main focus of the present paper are
the corrections due to the spin, we will here briefly outline how to obtain
the quantum correction due to particle dispersive effects. The general
evolution equation containing both spin and particle dispersive effects was
derived in Ref. \cite{Full-quantum}. Here we take that evolution equation,
drop the spin effects considered above, and keep the particle dispersive
effects in a weak quantum expansion (with the thermal de Broglie wavelength
over the characteristic scale length as expansion parameter), where only
terms up to $\hbar ^{2}$ are kept. The governing equation then reads 
\begin{align}
\frac{\partial }{\partial t}& f+\mathbf{v}\cdot \nabla _{x}f+\frac{q}{m}%
\left( \mathbf{E}+\mathbf{v}\times \mathbf{B}\right) \cdot \nabla _{v}f 
\notag \\
=& \frac{\hbar ^{2}}{24m^{2}}\left\{ \frac{q}{m}\left( \mathbf{E}+\mathbf{v}%
\times \mathbf{B}\right) \cdot \nabla _{v}\left( \overleftarrow{\nabla }%
_{x}\cdot \overrightarrow{\nabla }_{v}\right) ^{2}-2\left[ \frac{q}{m}%
\mathbf{B}\times \nabla _{v}\left( \overleftarrow{\nabla }_{x}\cdot 
\overrightarrow{\nabla }_{v}\right) \right] \cdot \left( \frac{q}{m}\mathbf{B%
}\times \nabla _{v}+\nabla _{x}\right) \right\} f
\end{align}%
Defining 
\begin{equation}
f=F_{0}+\tilde{f}
\end{equation}%
where $F_{0}$ is the background distribution and $\tilde{f}$ is the
perturbed distribution function, we can separate \eqref{vlasov} in linear
and nonlinear terms: 
\begin{align}
\frac{\partial }{\partial t}& f+\mathbf{v}\cdot \nabla _{x}f+\frac{q}{m}%
\left( \mathbf{E}+\mathbf{v}\times \mathbf{B}\right) \cdot \nabla _{v}F_{0}-%
\frac{\hbar ^{2}q}{24m^{3}}\left( \mathbf{E}+\mathbf{v}\times \mathbf{B}%
\right) \cdot \nabla _{v}\left( \overleftarrow{\nabla }_{x}\cdot 
\overrightarrow{\nabla }_{v}\right) ^{2}F_{0}  \notag \\
=& -\frac{q}{m}\left( \mathbf{E}+\mathbf{v}\times \mathbf{B}\right) \cdot
\nabla _{v}\tilde{f}+\frac{\hbar ^{2}}{24m^{2}}\bigg\{\frac{q}{m}\left( 
\mathbf{E}+\mathbf{v}\times \mathbf{B}\right) \cdot \nabla _{v}\left( 
\overleftarrow{\nabla }_{x}\cdot \overrightarrow{\nabla }_{v}\right) ^{2}%
\tilde{f}  \notag \\
& -2\left[ \frac{q}{m}\mathbf{B}\times \nabla _{v}\left( \overleftarrow{%
\nabla }_{x}\cdot \overrightarrow{\nabla }_{v}\right) \right] \cdot \left( 
\frac{q}{m}\mathbf{B}\times \nabla _{v}F_{0}+\nabla _{x}\tilde{f}\right) %
\bigg\}
\end{align}%
Assuming two waves as in previous calculations, and a Maxwellian background
distribution we have 
\begin{equation}
f_{1,2}\approx -\frac{iq(\mathbf{E}_{1,2}+\mathbf{v}\times \mathbf{B}%
_{1,2})\cdot \nabla _{v}\left[ 1-\frac{\hbar ^{2}}{24m^{2}}(i\mathbf{k}%
_{1,2}\cdot \overrightarrow{\nabla }_{v})^{2}\right] F_{0}}{m(\omega _{1,2}-%
\mathbf{k}_{1,2}\cdot \mathbf{v})}.
\end{equation}%
where we separated the perturbed distribution function into a linear part $%
f_{1,2}$ and a nonlinear part $f_{nl}$. The rest of the calculations can be
performed as in section III, although the algebra gets extremely complicated
in general. A thorough treatment of particle dispersive effects is beyond
the scope of the present paper, but we will nevertheless point out two
conclusions. Firstly, that the particle dispersive effects can be comparable
in magnitude to the spin contributions. Secondly, although these two quantum
effects may be comparable, they do not typically cancel, as the spin current
for parallel propagation parallel polarization scales as $\propto \hbar
^{2}k_{1}k_{2}B_{1}B_{2}$ (see eq. (\ref{par-pol.prop})), whereas the
corresponding scaling can be shown to be $\propto \hbar
^{2}B_{1}B_{2}/k_{1}k_{2}$ for particle dispersive effects. Thus we conclude
that the contribution of particle dispersive effects to the nonlinear
current density is worthy of consideration in its own right, but we should
not expect such contributions to cancel those due to the spin.

\end{document}